\PassOptionsToPackage{dvipsnames}{xcolor}
\documentclass[twocolumn,times,twocolappendix]{aastex63}

\usepackage{savesym}
\savesymbol{tablenum}
\usepackage{siunitx}
\restoresymbol{SIX}{tablenum}
\usepackage{amsmath}
\usepackage{CJK}
\usepackage{bm}
\usepackage[normalem]{ulem}
\usepackage{soul}
\usepackage{enumitem}

\usepackage{pgfgantt}

\usepackage{newunicodechar,graphicx}

\DeclareRobustCommand{\okina}{%
  \raisebox{\dimexpr\fontcharht\font`A-\height}{%
    \scalebox{0.8}{`}%
  }%
}
\newunicodechar{ʻ}{\okina}

\begin{document}

\begin{CJK*}{UTF8}{gbsn}

\title{Refinements to the Solar Polar Magnetic Flux: Implications from Inversion Methodologies}

\author[0000-0002-0198-0528]{Bryan Yamashiro} 
\altaffiliation{\textit{DKIST} Ambassador}
\affil{Institute for Astronomy, University of Hawai$\okina$i at M\={a}noa, 2680 Woodlawn Drive, Honolulu, HI 96822, USA}

\author[0000-0003-4043-616X]{Xudong Sun (孙旭东)}
\affil{Institute for Astronomy, University of Hawai$\okina$i at M\={a}noa, 34 Ohia Ku Street, Pukalani, HI 96768, USA}

\author[0000-0002-0189-5550]{Ivan Mili\'{c}}
\affil{Institute for Solar Physics, Georges-K\"{o}hler-Allee 401a
79110 Freiburg, Germany}

\author[0000-0001-5518-8782]{Carlos Quintero Noda}
\affil{Instituto de Astrof\'{i}sica de Canarias, E-38205, La Laguna, Tenerife, Spain}
\affil{Departamento de Astrof\'{i}sica, Univ. de La Laguna, La Laguna, Tenerife, E-38200, Spain}

\author[0000-0003-2359-9039]{Adur Pastor Yabar}
\affil{Institute for Solar Physics, Department of Astronomy, Stockholm University, AlbaNova University Centre, 10691 Stockholm, Sweden}

\author[0000-0002-1327-1278]{Rebecca Centeno}
\affil{High Altitude Observatory (NSF NCAR), 3080 Center Green Dr., Boulder, CO, 80301, USA}

\author[0000-0002-7290-0863]{Jiayi Liu}
\altaffiliation{\textit{DKIST} Ambassador}
\affil{Institute for Astronomy, University of Hawai$\okina$i at M\={a}noa, 2680 Woodlawn Drive, Honolulu, HI 96822, USA}

\author[0000-0002-5879-4371]{Milan Go\v{s}i\'{c}}
\affil{Lockheed Martin Solar and Astrophysics Laboratory, Palo Alto, CA 94304, USA}
\affil{SETI Institute, Mountain View, CA 94043, USA}

\author[0000-0002-7663-7652]{Kai Yang}
\affil{Institute for Astronomy, University of Hawai$\okina$i at M\={a}noa, 34 Ohia Ku Street, Pukalani, HI 96768, USA}


\begin{abstract}
The magnetic fields in the solar polar region are important to our understanding of the internal dynamo process, the global coronal structure, and the origin of the solar wind. The inference of polar fields based on spectropolarimetric observation is highly model-dependent and can suffer from various systematic effects. Here we analyze a raster map of the southern polar region taken by the \textit{Hinode} Spectro-Polarimeter, utilizing the Stokes Inversion based on Response functions code. The inversions provide height-dependent vector magnetic field maps between optical depths $\log_{10}\tau = -2$ and $0$. We examine the impact on the total magnetic flux estimate from adopting (1) 1- vs 2-component atmospheric models via a ``filling factor'' parameter and (2) different analysis schemes. At $\log_{10}\tau = -1.5$, the polar magnetic flux is estimated to be $(1.84 \pm 0.03) \times 10^{21}$~\si{Mx} and $(1.38 \pm 0.02) \times 10^{21}$~\si{Mx} under the 1- and 2-component atmosphere assumption, respectively. The magnetic flux is approximately constant or increases slightly with height, respectively. We find that the 2-component (1-component) configuration is preferred for 58.3\%\,(32.3\%) of the pixels. Different initial guesses, including the input atmosphere model and the filling factor, as well as different inversion settings, can significantly affect the results, especially for locations with weaker polarization signals. Our work highlights the importance of including unresolved magnetic structures or stray light into consideration. Model degeneracy and the convergence to local minima limit the precision of the polar magnetic flux inference (no better than several tens of percent in this case). Higher-resolution observations and advanced inversion and disambiguation algorithms may alleviate these limitations.
\end{abstract}




\section{Introduction} \label{sec:introduction}

The Sun's magnetic fields in the polar regions (polar fields) contribute substantially to the overall topology of the global-scale magnetic field, the coronal plasma structure, and the origin of the solar wind. During the minimum phase of a solar cycle, the polar fields are mostly unipolar and open, forming large polar coronal holes of lower density and temperature with respect to the surroundings. They appear dark in most extreme ultraviolet and X-ray images and are {a} source of the fast solar wind. Polar fields during these times account for most of the magnetic flux in the heliosphere, dictate the heliospheric structure, and influence the activities in interplanetary space \citep{thompson2011}. The mean polar magnetic fields also embody the interior poloidal fields that are the progenitor of the new-cycle toroidal field and correlate well with the sunspot number of the upcoming maximum phase \citep{dikpati2007}.

Observations of the solar polar regions, typically latitudes beyond $\pm 60^{\circ}$, are hindered by Earth's ecliptic viewpoint. The \ang{7.25} tilt of the solar rotation axis with respect to the ecliptic inherently specifies optimal viewing angles at Earth, in early March for the south pole and early September for the north pole. Observations are also affected by strong intensity gradients and foreshortening at the solar limb, and additionally by variable seeing conditions for ground-based instruments \citep{petrie2015}.

During the solar minima, open magnetic flux derived from photospheric maps is expected to match that derived from \textit{in situ} observations. However, a recent study found that the former is underestimated compared to the latter by a factor of two or more \citep{linker2017}. This ``open flux problem'' may result from a variety of factors \citep[see][and references therein]{arge2024}. One line of argument focuses on the unknown biases in remote sensing observation or analysis methods, further discussed in Section~\ref{subsec:literature}. It is also one of our motivations and will be further explored in this work.

Measurements of the polar fields traditionally probe the line-of-sight (LOS) component at lower resolution. Following the introduction of the solar magnetograph \citep{babcock1953} and initial observations of the polar regions \citep{babcock1959}, daily observations of the Sun's LOS magnetic fields were notably conducted by the Mount Wilson Solar Observatory \citep{howard1989}. Early studies noted the unipolar polarities of polar fields \citep{babcock1955} and speculated that they are comprised of tiny magnetic elements corresponding to the predominant polarity \citep{gillespie1973}. 
Higher-resolution magnetic field observations unveiled the unipolar elements to be ubiquitous in the polar landscape. Using the video magnetograph of the Big Bear Solar Observatory ($\sim0\farcs6$ resolution), \cite{lin1994} showed that the polar region magnetic flux is indeed concentrated in small magnetic elements. With space-based observations from the \textit{Hinode} satellite \citep{kosugi2007}, \cite{tsuneta2008} demonstrated that most polar magnetic flux is concentrated in small patches ($\sim$\ang{;;1} to \ang{;;5}) in the photosphere which occupies a small fraction of the polar region. Subsequent polar field studies \citep[e.g.,][]{ito2010,pastoryabar2020} confirmed the presence of large flux concentrations exceeding $10^{16}$~\si{Mx} within polar coronal holes.

A quantitative evaluation of the polar magnetic field requires spectropolarimetric inversion. The procedure infers the magnetic field and other physical parameters by iteratively solving for the radiative equation. It is generally an ill-posed problem and can be particularly challenging for polar observations. For example, a large fraction of the polar fields are assumed to be perpendicular to the solar surface \citep{svalgaard1978,ito2010}, i.e., transverse to the LOS at high latitudes. The relevant signal thus appears mainly as linear polarization (Stokes $Q$ and $U$), which is inherently less sensitive compared to circular polarization (Stokes $V$) and thus more susceptible to noise. Moreover, the foreshortening effect at high latitudes degrades effective spatial resolution, blending unresolved structures, and further suppressing the polarization signal through cancellation. Finally, the vector field direction cannot be fully resolved by inversion alone due to the $180^\circ$ azimuthal ambiguity intrinsic to the Zeeman effect. The weaker linear polarization signal, coupled with the foreshortening effect due to the inclined viewing angle, can make the disambiguation task challenging. Nevertheless, disambiguation is necessary to quantify the amount of open field, as the ambiguity exists in the observer's frame, while we aim to probe the local frame.

Recent studies highlight the limitations of existing techniques. \cite{sun2021} finds that inferences made with variable filling factor $f$ (used to describe unresolved structure or stray light, see Sec.~\ref{subsubsec:modelatmos}) incur less inclined and more radial fields compared to the $f\equiv1$ counterpart, but the trend may be a bias inherent to limited resolution. \cite{centeno2023} showed that Milne-Eddington (ME) atmospheres (often adopted for inversion, e.g. in \citealt{tsuneta2008}) with variable $f$ cannot adequately model the spatial mixing that arises from the polar field LOS foreshortening and the telescope point spread function (PSF). For lower-resolution observations of weaker magnetic fields (outside of sunspots), systematic biases of the transverse field, as either over- or under-estimates, were found in the inversion results \citep{pevtsov2021,leka2022}. Detailed analysis procedures were found to be important in mitigating biases.

In this work, we explore the inference of polar magnetic fields using high-resolution \textit{Hinode}/SP spectropolarimetric data. In particular, we will examine the impact of adopting two-component atmospheric models via filling factor. We also discuss the additional parameter degeneracies brought by this approach.
Section~\ref{sec:methods} below describes the data and details of the inference procedures. Section~\ref{sec:results} presents the resultant magnetic flux of the polar region and its dependence on various modeling assumptions. Section~\ref{sec:discussion} compares our result to the literature, including inversion of the same dataset from a public database \citep{lev2}, and discusses possible biases and the implications of our work.


\section{Data and Methods} \label{sec:methods}

\begin{figure*}[t!]
\centering
\includegraphics[width=\textwidth]{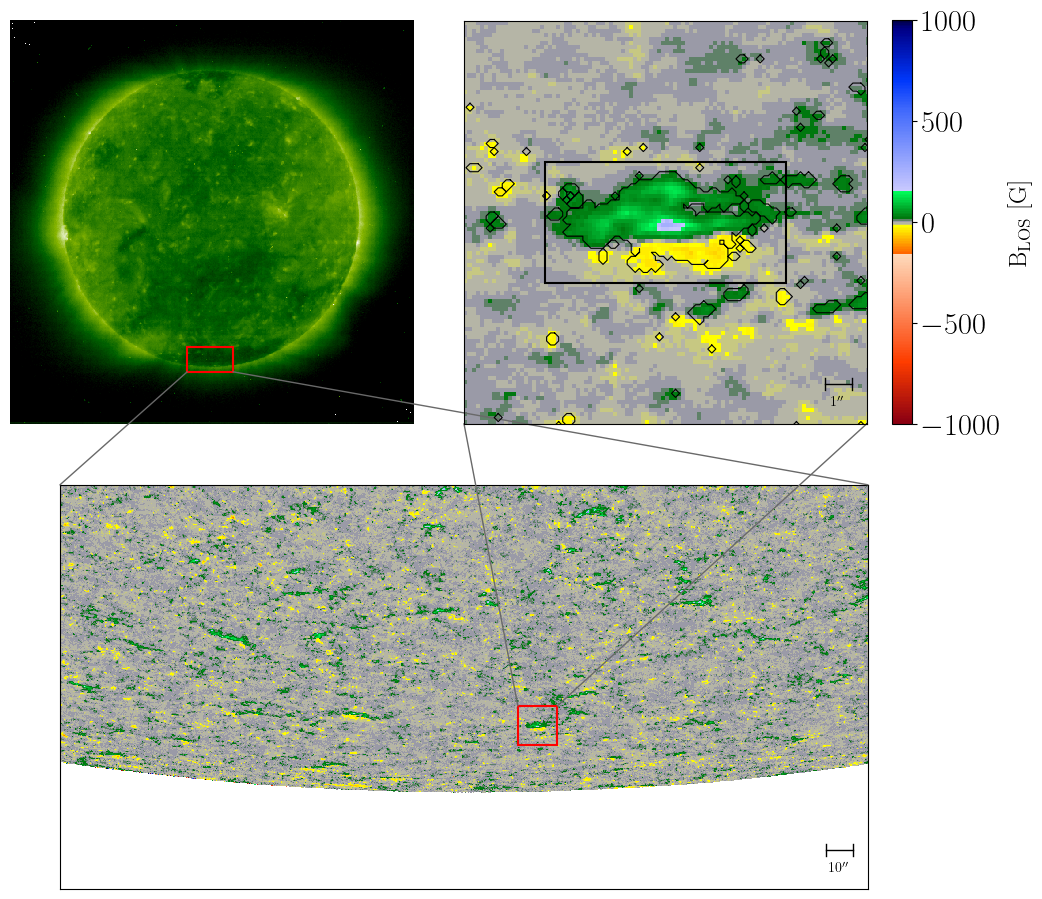}
\caption{\textit{Top Left:} Full disk \SI{195}{\AA} \textit{SOHO}/EIT EUV image taken on 2007.03.16. \textit{Bottom:} \textit{Hinode}/SP $B_{\mathrm{LOS}}$ map of the southern polar coronal hole. The field of view is shown by the red box in the top left panel. \textit{Top Right:} Zoomed view of a polar magnetic patch as indicated by the red box in the bottom panel. The pixels with significant polarization selected for analysis are outlined by black contours. The black box is the window used for the height analysis in Section~\ref{subsec:height}.}
\label{polar_field_context}
\end{figure*}

We focus on a southern polar coronal hole observed on 2007 March 16, shown in Figure~\ref{polar_field_context}, by the Spectro-Polarimeter \citep[SP;][]{lites2013} instrument of the Solar Optical Telescope \citep[SOT;][]{tsuneta2008b} onboard \textit{Hinode} \citep{kosugi2007}. SP is a slit scanning spectro-polarimeter with a pixel scale along the slit of $0\farcs16$. The instrument observes the two Zeeman-sensitive \ion{Fe}{1} lines at the 630~nm window.
Here, the raster scan covered a field of view (FOV) of $327\farcs52$ (east-west) by $163\farcs84$ (north-south). The south pole was tilted toward Earth, which provides a favorable vantage point to study the polar fields. The top left panel of Figure~\ref{polar_field_context} depicts the raster area (red box) with respect to 
a full disk \SI{195}{\AA} EUV image taken by the Extreme ultraviolet Imaging Telescope \citep[EIT;][]{delaboudiniere1995} from the \textit{Solar and Heliospheric Observatory} \citep[\textit{SOHO};][]{domingo1995}. The observation was made during the descending solar minimum of Solar Cycle 23. The bottom panel displays a reference \textit{Hinode}/SP LOS field $B_{\mathrm{LOS}}$ raster of the southern polar coronal hole (see Section~\ref{sec:sir}). Strong field patches are mostly positive in $B_{\mathrm{LOS}}$, coinciding with the overall positive polarity of the south pole. Part of the negative polarity in $B_{\mathrm{LOS}}$ is due to projection. We will refer to this FOV as the ``full map" henceforth. The top right panel is a zoomed view of a polar magnetic patch containing significant polarization, as indicated by the red box in the bottom panel. Pixels with significant polarization selected for analysis are outlined by black contours. The black box is a $60\times30$~px sub area of the full map, referred to as the ``window" FOV, used for the height analysis in Section~\ref{subsec:height}.


\subsection{Limb Darkening Correction}

Inversion algorithms typically require normalization of the Stokes profiles, taken to be the mean intensity of the quiet-Sun continuum $I_c$ at the disk center. Because $I_c$ decreases rapidly and non-linearly toward the limb, an additional, spatially dependent extrapolation is required to properly account for limb darkening. This procedure provides a theoretical disk center continuum normalization for pixels closer to the limb. We adopt the limb darkening relation from \cite{pierce1977}, $I_{\lambda}(\mu) / I_{\lambda}(1)$, for wavelength $\lambda=6326$~\AA~(the closest available values in wavelength to the \ion{Fe}{1} lines). Here, $\mu$ is the cosine of the angle between the surface normal and the observer's line of sight, where the disk center is $\mu=1$. $I_{\lambda}(\mu)$ is the continuum intensity.  For this SP raster, $0\le\mu\le0.45$, the observed continuum intensity for a single pixel is defined as the mean value of the Stokes $I$ profiles between the two \ion{Fe}{1} lines.

\begin{figure}
\centering
\includegraphics[width=0.45\textwidth]{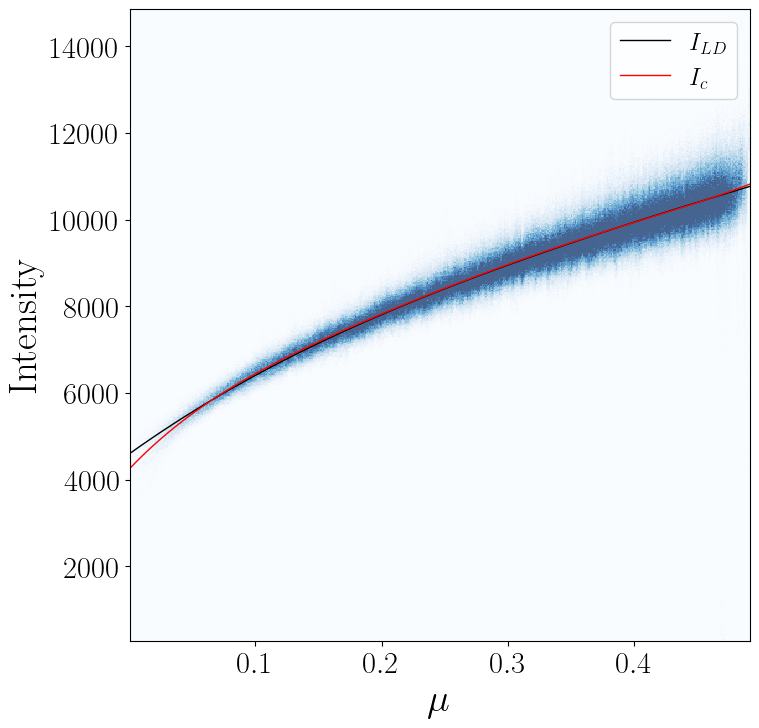}
\caption{Continuum intensities of the \textit{Hinode}/SP raster. The blue point cloud shows the values of all pixels. The mean intensity $I_c$ as a function of $\mu$ is represented by the red line. The black line $I_{LD}$ depicts the reference limb darkening curve multiplied by the scaling term that minimizes the squared error between the two curves. This curve is used for normalizing the observations.}
\label{LD_ic_distribution}
\end{figure}

The blue points in Figure~\ref{LD_ic_distribution} show the $I_c$ distribution as a function of $\mu$ for the {full map}. Taking the mean $I_c$ with respect to $\mu$, shown in red, and the theoretical limb darkening relation, we devise a scaling that matches one curve to the other by using the Nelder-Mead optimization to minimize the squared difference between the two curves (black curve). All subsequent Stokes spectra are divided by this scaling term.

\subsection{Pixel Selection} \label{subsec:pixelselection}


Only pixels containing strong polarization signals are selected for inversions. We first estimate the polarimetric noise $\sigma$ from the standard deviations of the normalized Stokes $Q,U$ and $V$ profiles in the continuum between the two Fe\,I lines. For this specific dataset the noise is approximately $\sigma=2\times 10^{-3}$. Then, we select only the pixels where at least five wavelength points exhibit polarization amplitudes, in any of the Stokes parameters, above $5\sigma$. Additionally, we require at least three of these wavelength points to be consecutive. This selection is much more rigorous than the one in the earlier works \citep[e.g.][]{orozcosuarez2007,tsuneta2008,ito2010}, but we found that it helps ensure that the selected pixels have high enough polarization to enable meaningful spectropolarimetric inversion.

Additionally, we require the inverted inclination values to be within $[0^\circ,180^\circ]$, as the interpolation in the inversion code (between the pre-defined ``nodes'' in $\tau$, see below) may introduce non-physical negative values. In total, 3.6\% of the pixels are considered significant and the remainder of the analysis utilize these as the ``{full map mask}.'' On average, the selected magnetic patches span $0\farcs85$, the largest extending up to $16\farcs19$.

We note that our stringent selection criteria are expected to yield a lower estimate of the magnetic flux compared to that of the literature. We discuss the implications in Section~\ref{sec:discussion}.


\begin{deluxetable}{cccc}[t]
\tablecaption{SIR configuration parameters. \label{tab:mathmode}}
\tablecolumns{3}
\tablewidth{0pt}
\tablehead{\colhead{Quantity}  & \colhead{Component 1} & \colhead{Component 2} }
\startdata
 Temperature ($T$)           & 2,3,5      & 2,3,3  \\
 Microturbulence ($v_{\mathrm{micro}}$)       & 1,1,2      & 0 (1)\tablenotemark{a}  \\
 Field Strength ($B$)       & 1,2,3      & 0   \\
 LOS Velocity ($v_{\mathrm{LOS}}$)         & 1,2,3      & 1,2,2   \\
 Inclination ($\gamma$)                 & 1,2,3      & 0  \\
 Azimuth ($\psi$)                   & 1,1,1      & 0  \\ \hline
\enddata
\tablecomments{Component 1 and 2 are for the magnetic and non-magnetic atmospheric component, respectively. The listed values are the number of nodes set for the specified quantity for three cycles of iteration.}
\tablenotetext{a}{Microturbulence is included as a height-independent free parameter for the non-magnetic component, for one special run discussed in Section~\ref{subsec:effect2}.}
\label{table:sirparams}
\end{deluxetable}


\subsection{Spectropolarimetric Inversion} \label{sec:sir}

Inversions are performed for the selected strong field pixel profiles with the Stokes Inversion based on Response functions \citep[SIR;][]{ruizcobo1992}. SIR is capable of synthesis and inversions of spectral lines formed in the presence of magnetic fields, under the assumption of local thermodynamic equilibrium (LTE), accounting for Zeeman-induced polarization.

The inversion module minimizes the reduced chi-square merit function ($\chi_\mathrm{red}^2$) which is defined from \cite{ruizcobo1992} as:
\begin{equation}
    \chi_\text{red}^2 \equiv \frac{1}{\nu} \sum_{k=1}^{4} \sum_{i=1}^{N} \left [I_k^{\text{obs}}(\lambda_i) - I_k^{\text{syn}}(\lambda_i)  \right]^2 \frac{w_{ki}^2}{\sigma_{ki}^2}.
    \label{chir}
\end{equation}

Here, the Stokes vectors are represented by the indices $k=1,...,4$, sampled at wavelengths $i=1,...,N$. The remaining parameters are the degrees of freedom $\nu$, uncertainties of the observations $\sigma_{ki}$, weighting factors $\omega_{ki}$, and the observed ($I_{k}^\text{obs}$) and modeled ($I_{k}^\text{syn}$) Stokes profiles. The $\omega_{ki}$ are set to ($\omega_I=1$, $\omega_Q=5$, $\omega_U=5$, $\omega_V=2$) assigning more weight to the linear polarization because it is generally weaker but more important to the polar magnetic flux estimate. For this observation, SP has a total of $N=112$ observed wavelengths per spectra.

\subsubsection{Model Atmosphere}

The SIR algorithm allows the physical parameters to vary along the optical depth $\tau$ and can invert multiple spectral lines at the same time. In the case of the \ion{Fe}{1} 630~nm doublet, the Stokes $V$ response functions exhibit sensitivity to magnetic fields from $\log_{10}\tau$ = $0$ to $-2$ \citep{quinteronoda2021}. A model atmosphere used by SIR consists of $\tau$-dependent physical parameters: temperature $T$, line-of-sight velocity $v_{\rm los}$, microturbulent velocity $v_{\rm turb}$, and the magnetic field strength $B$, inclination $\gamma$ and azimuth $\psi$.

The stratification in optical depth is achieved using \textit{nodes} in the $\log_{10}\tau$ space for the specified atmospheric quantity. Nodes indicate the optical depth locations where SIR modifies the initial atmosphere to match the observed spectra, employing a Marquardt algorithm \citep{press1986} to minimize the differences between $I_k^{\text{obs}}$ and $I_k^{\text{syn}}$. This process is performed iteratively through several ``cycles'' (three in this study). The number of nodes is pre-determined for each cycle and is typically increasing with cycle, to provide the optimum complexity of the final atmosphere.
The nodes are placed evenly in the $\log_{10}\tau$ space specified by the initial atmosphere and the values between the nodes are interpolated. The node settings that we use are listed in Table~\ref{table:sirparams}. $Q$ and $U$ polarization signals are relatively weak and noisy in the polar region, and are expected to provide little constraint on the variation of azimuth with $\tau$. We therefore restrict the fitting of azimuth to one node to avoid spurious results from the inversion.

The specific choice of the initial values of atmospheric parameters is known to significantly influence the inversion results, because of the inversion degeneracies and photon noise \citep[e.g.][]{MJMG_2006_validity}. To mitigate this effect, we use a set of 12 model atmospheres as the initial guesses. These were created by adding a range of different magnetic field strengths and orientations, as well as different LOS velocities to the semi-empirical models presented in \cite{gingerich1971} and \cite{fontenla1993}. The initial guesses for LOS velocity and magnetic field are constant in height (see the relevant three middle panels of Fig.\,\ref{modelatmospheres}). 

\subsubsection{Two-Component Model Atmosphere} \label{subsubsec:modelatmos}



We use a 2-atmospheric component configuration, one magnetic and one non-magnetic, with an adjustable filling factor parameter $f$ that specifies the fraction of the particular pixel occupied by the magnetic component. SIR assumes $f$ is constant with optical depth. Note that only the magnetic atmosphere contributes to the polarization. In this work, SIR infers both the magnetic and the non-magnetic atmospheric components, as well as the filling factor, to find the combination of atmospheres that best fit the data. The non-magnetic atmosphere is initialized with identical initial guess atmospheric parameters as the magnetic counterpart, but with the magnetic field set to zero. We use a default initial filling factor set to $0.5$.

\begin{figure*}
\centering
\includegraphics[width=0.9\textwidth]{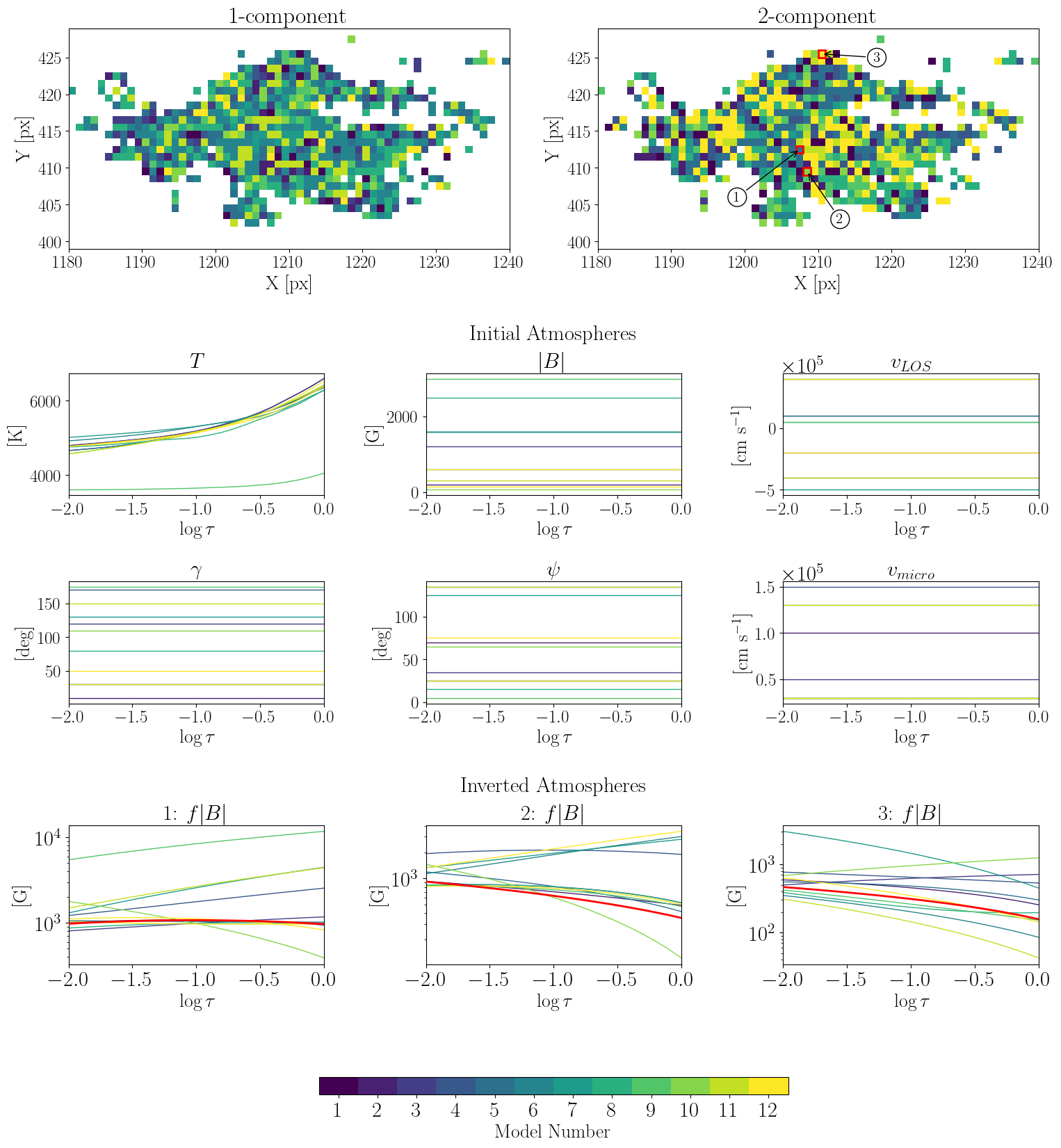}
\caption{The model number (1 to 12) of the initial model atmospheres that yield the lowest $\chi_\text{red}^2$ for each pixel, for the 1- (\textit{top left}) and 2- (\textit{top right}) component configurations. This has the same FOV as the zoomed window region in Figure~\ref{polar_field_context}. The bottom two rows show the profiles of each input atmosphere. The bottom row shows the inverted $f|B|$ for the three pixels indicated in the top right panel for the respective input atmosphere. The lowest $\chi_{\mathrm{red}}^2$ solutions are depicted with red lines.}
\label{modelatmospheres}
\end{figure*}

The top panels of Figure~\ref{modelatmospheres} show the initial model atmospheres that provide the best inversion, with the lowest $\chi_\text{red}^2$ value, for the 1- and 2-component configurations, respectively. No single initial guess consistently yields better results than the rest. The middle panels show the parameters that constitute each initial atmosphere.

The bottom panels illustrate the spread of 2-component $f|B|$ solutions that result from different initial model atmospheres, for the three pixels indicated in the top right panel. The lowest $\chi_{\mathrm{red}}^2$ allows us to determine the initial model that leads to the best-fit solution (red curves), model selection is further explored in Sec.\,\ref{subsec:modeldegeneracy}. The three pixels 1, 2, and 3 represent where the final $f|B_r|$ is the highest (1) and the lowest (2) in the small window, as well as at the edge of the patch (3). At $\log_{10}\tau = -1.5$, the differences between the extremes of $f|B|$ are rather substantial. Results that deviate significantly from the best-fit solution often correspond to much greater $\chi_{\mathrm{red}}^2$. They are likely converging to local minima in the $\chi_{\mathrm{red}}^2$ space. 

In this setup, the number of total free parameters, $p$, is found by summing the maximum nodes for each atmospheric parameter in the last inversion cycle, which is $p=17$ for the 1-component configuration. The 2-component configuration, on top of the five additional nodes for the non-magnetic atmospheric portion, includes an additional free parameter for the filling factor, resulting in $p=23$. Degrees of freedom are defined as the number of observables minus the number of free parameters for that configuration, $\nu = 4N-p$, where the additional factor of 4 accounts for the four Stokes profiles.

SIR estimates and outputs the errors associated with each inverted atmospheric quantity under the assumption that the model parameters are independent \citep{ruizcobo1992}. Uncertainties are proportional to the inverse of the response functions to changes in the physical quantity during the inversion process.

\subsubsection{Instrumental Effects}


Stray light, typically defined as the non-local contribution to the Stokes spectrum of the analyzed pixel, is known to influence spectropolarimetric inversions \citep[][]{labonte2004, orozcosuarez2007}. Several studies have adopted a local treatment of stray light, in which each pixel is corrected using a spatially dependent contribution from its neighborhood \citep[e.g.,][]{asensio2011}. Other works instead employ a global correction, applying a single average stray-light profile to all pixels \citep[e.g.,][]{orozcosuarez2007}. Both approaches rest on assumptions that may introduce their own systematic biases. The effect should be largely mitigated by using spatially-coupled inversion that explicitly take into account the spatial point spread function of \textit{Hinode}/SOT \citep[][]{vanNoort_2012_speccoupl}, but this approach has not been tested for observations close to the limb. Here, we do not introduce an explicit stray-light component into the inversions, but explore a two-component model atmosphere, where the non-magnetic component, in a way, mimics the effect of the stray light.

SIR accounts for instrumental broadening by convolving the forward-modeled Stokes profiles with the \textit{Hinode}/SP spectral point spread function (PSF) derived from the optical characterization \citep{lites2013}. Scattered light (constant offset to the Stokes $I$ due to the scattering of the light in the spectrograph) is not accounted for, as we assume it does not significantly influence the inference of the magnetic fields.


\begin{figure*}
\centering
\includegraphics[width=1.0\textwidth]{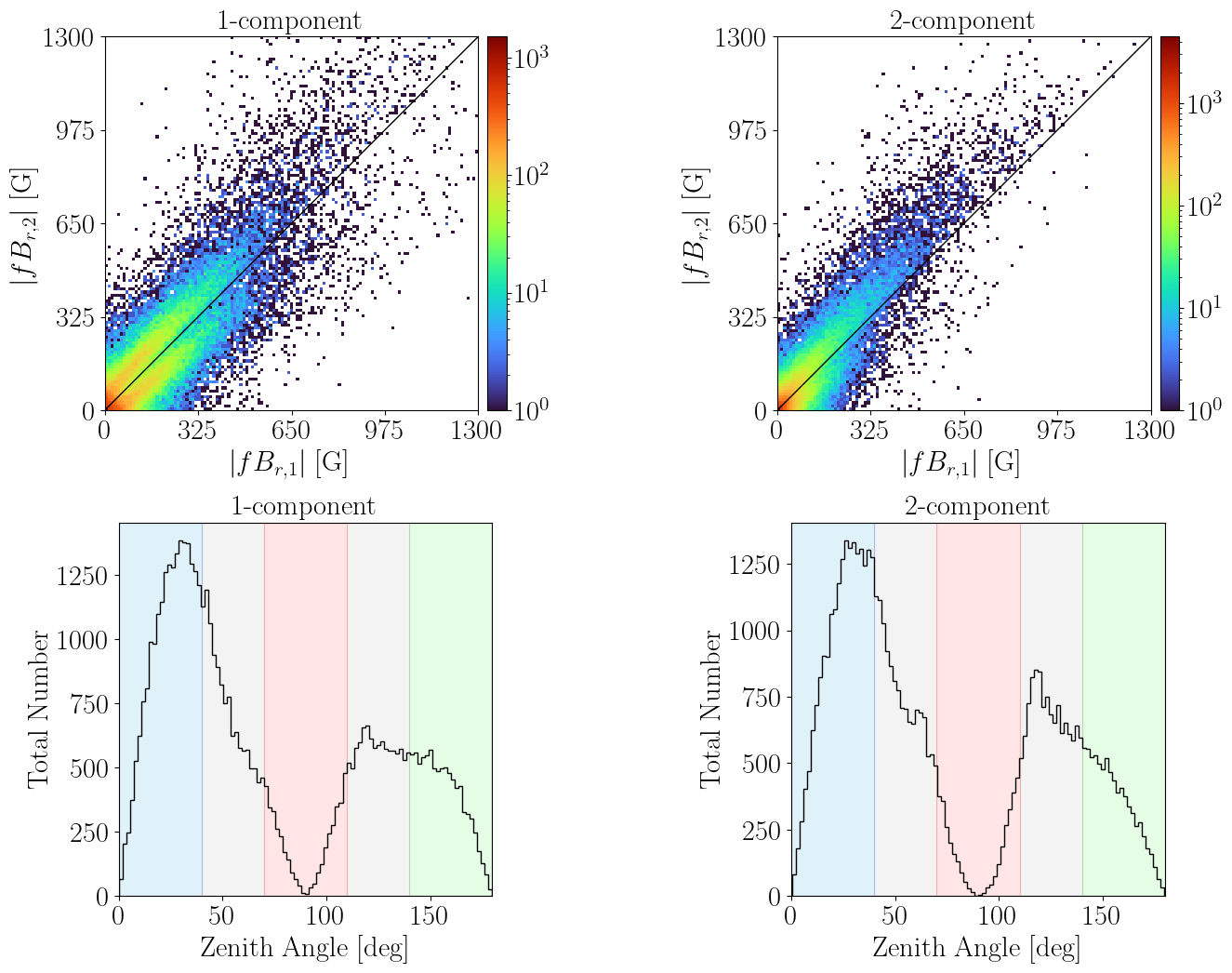}
\caption{\textit{Top Panels:} Distributions of $|fB_{r,1}|$ and $|fB_{r,2}|$ for the 1-component (left) and 2-component (right) configurations at $\log_{10}\tau = -1.5$. \textit{Bottom Panels:} Zenith angles for the magnetic field with respect to the local normal, 1-component (left) and 2-component (right) configurations at $\log_{10}\tau = -1.5$. The shaded regions indicate where the field is considered vertical (blue, green), horizontal (red), and undefined (gray).}
\label{i1i2hist}
\end{figure*}


\subsection{Azimuth Disambiguation and Vector Transformation}\label{subsec:disambiguation}

The azimuth value derived from Zeeman-based inversions suffers from an intrinsic \ang{180} ambiguity in the LOS frame. Resolving the ambiguity is challenging in the polar region, but is crucial to the magnetic flux estimate because the radial field $B_r$ largely depends on the transverse field component. Here, we adopt the simple radial-acute method by picking the solution that maximizes the absolute value of the radial magnetic field component $|B_r|$. This approach is justified in coronal holes, where the photospheric field is expected to be predominantly vertical and concentrated in unipolar flux structures, with weaker and more disorganized horizontal fields \citep{petrie2009, shiota2012}.

The inverted magnetic field vector is output as the field strength $B$, inclination $\gamma$, and azimuth $\psi$, defined with respect to the LOS and the plane of sky. The azimuth angle is defined where $Q$ is maximized and $U = 0$, increasing counterclockwise with respect to the observer. To resolve the azimuthal ambiguity, we transform the field vector into a Heliocentric spherical coordinate as $(B_r,B_\theta,B_\phi)$ following \cite{gary1990} and \cite{sun2013}. The two potential azimuth solutions, $\psi_1=\psi$ and $\psi_2=\psi + \pi$, lead to two radial field values $B_{r,1}$ and $B_{r,2}$.

We choose the solution with the greater radial field magnitude and repeat the procedure for all optical depths independently. The top panels in Figure~\ref{i1i2hist} display the distributions of $|fB_{r,1}|$ and $|fB_{r,2}|$ at $\log_{10}\tau = -1.5$, respectively. About $33\%\,(46\%)$ of pixels adopt the $|fB_{r,1}|$ solution for the 2-component (1-component) configurations, and the rest adopt the $|fB_{r,2}|$ solution. The solutions for the 2-component (1-component) configurations disagree for a non-negligible amount of pixels.

The bottom panels of Figure~\ref{i1i2hist} show the distribution of the zenith angles (inclination with respect to the local normal) at $\log_{10}\tau = -1.5$ for the full FOV. Field orientations are classified as vertical ($0^\circ - 40^\circ$, $140^\circ - 180^\circ$; blue, green spans) or horizontal ($70^\circ - 110^\circ$; red span) to the local surface, or undefined (gray spans) following \citet{ito2010}. A total of $50.3\%\,(53.2\%)$ are vertical, $6.6\%\,(7.9\%)$ are horizontal, and $43.1\%\,(38.8\%)$ are undefined. Since our strategy selects the solution that results in making the vector field as close to the radial direction as possible, it is not surprising that most zenith angles are dominantly vertical.

We have also experimented with the ``minimum energy'' method \citep{metcalf1994,leka2009} and another geometry-based method \citep{ito2010}. A brief discussion can be found in Appendix~\ref{sec:appendix}.

After the disambiguation, the total magnetic flux $\Phi$ is computed by summing all inverted pixels:
\begin{equation}
    \Phi = \sum \frac{ B_r \times f \times dA}{\mu},
\end{equation}
where $dA$ is the physical area for a single pixel on the Sun and $f$ is the filling factor.

We estimate the uncertainty in $B_r$ using the error reported by SIR for $(B,\gamma,\psi)$ and the standard error propagation procedure following \cite{sun2013}. We ignore the correlations between these variables, as well as the error in the filling factor, as SIR currently does not report them. We finally estimate the uncertainty in $\Phi$ assuming that all pixels are independent.

\subsection{{Reference} Milne-Eddington Inversions}

We additionally obtain reference inversion results from the \textit{Hinode}/SP Level 2 dataset at CSAC/HAO \citep{lites2013b,lev2}. The dataset is processed with the Milne-Eddington gRid Linear Inversion Network \citep[MERLIN;][]{lites2007} algorithm, which is an ME inversion code based on the least-squares fitting of the observed Stokes profiles using the Levenberg-Marquardt algorithm \citep{skumanich1987}. Profiles are fit for $\pm0.3$\,\si{\AA} from the respective \ion{Fe}{1} line centers. The resultant magnetic field vectors and filling factors are compared with the SIR results. At this moment, the publicly available MERLIN data do not include reliable error estimates.


\section{Results}\label{sec:results}

\begin{figure}[t!]
\centering
\includegraphics[width=\columnwidth]{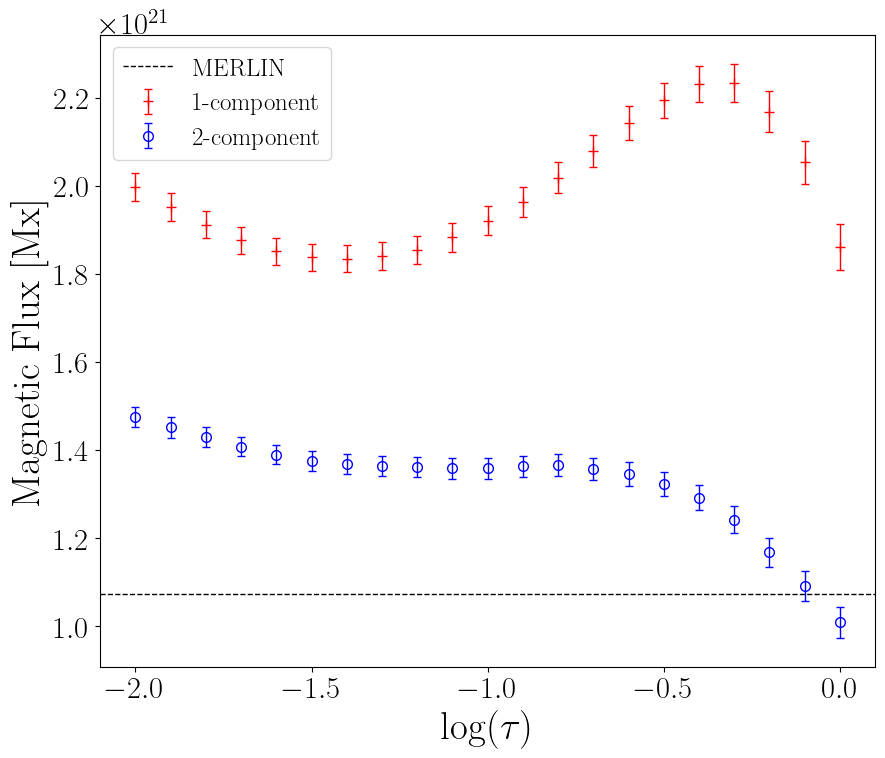}
\caption{Total polar magnetic flux of pixels containing significant polarization signals for the 1-component (red) and 2-component (blue) configurations. The dashed horizontal line represents the total flux calculated from the MERLIN inversion with our selected pixels.}
\label{btotflux}
\end{figure}

\begin{figure*}[t!]
\centering
\includegraphics[width=1.0\textwidth]{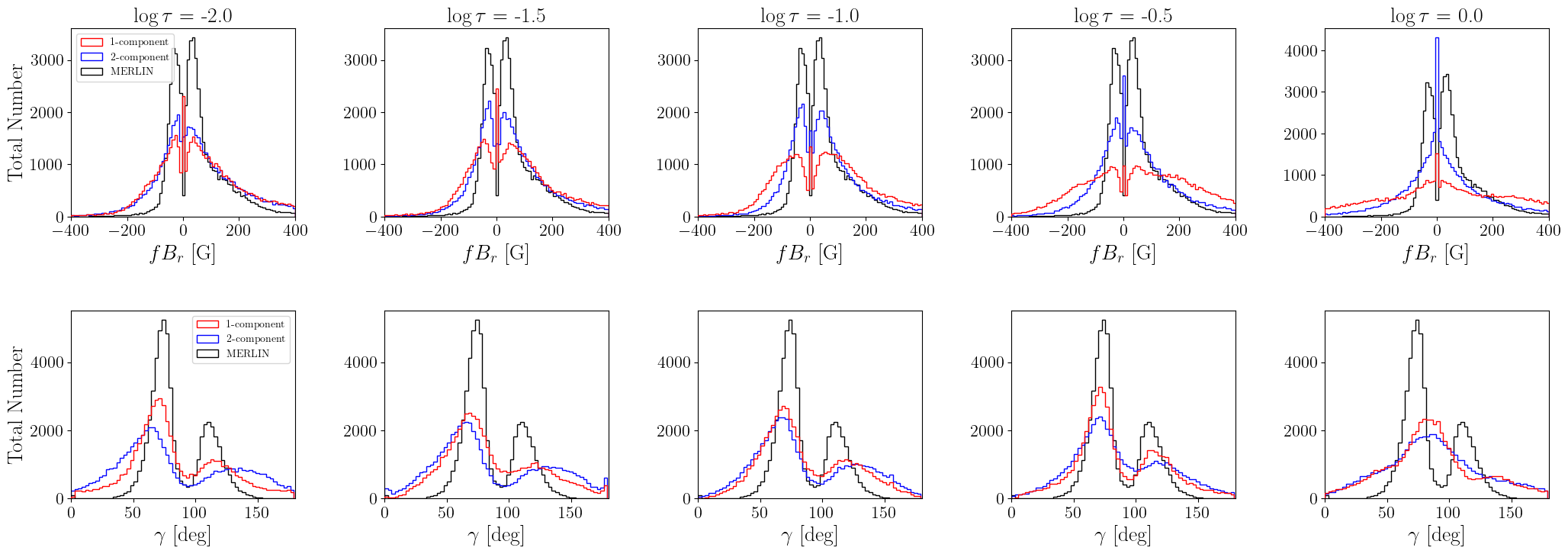}
\caption{Comparisons of the full map inversion results of the radial magnetic field $fB_r$ (top row) and inclination (bottom row) between the 1- (red) and 2-component (blue) configurations, for five different optical depth layers. MERLIN values (black) are added for reference and remain unchanged through the optical depth range.}
\label{model_compare}
\end{figure*}

\begin{figure*}[t!]
\centering
\includegraphics[width=\textwidth]{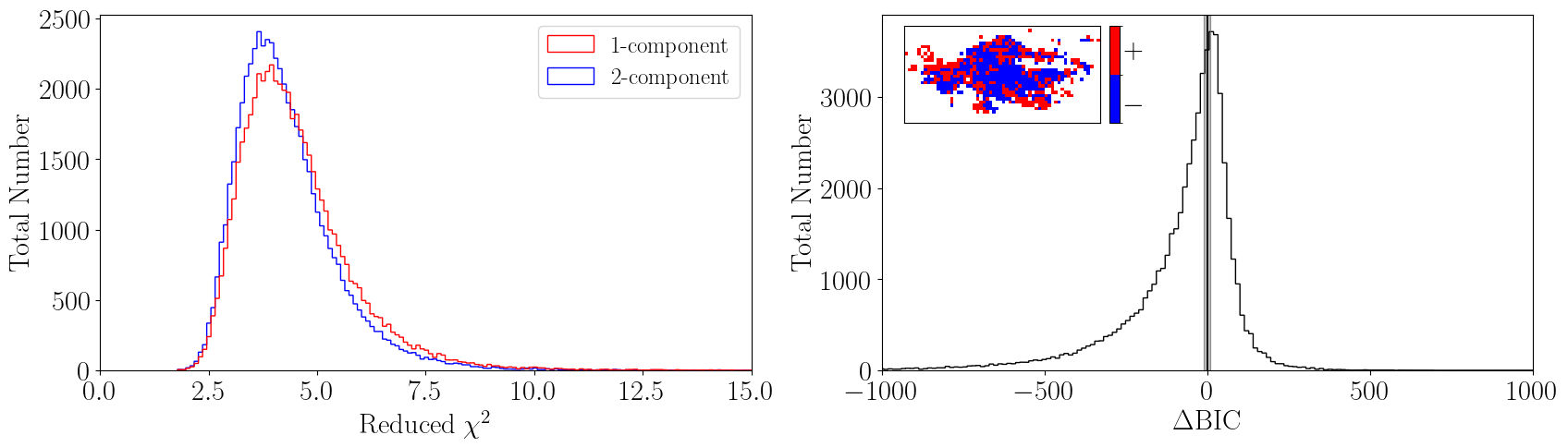}
\caption{\textit{Left:} Distribution of $\chi_\text{red}^2$ for the full raster map, for the 1-component (\textit{red}) and 2-component (\textit{blue}) model configurations. \textit{Right:} Difference of BIC values (see main text) between the 1- and 2-component models. $\Delta$BIC$=0$ is indicated by the vertical black line and $-10 \leq \Delta\textnormal{BIC} \leq 10$ by the gray span. The figure inset represents positive (\textit{red}) and negative (\textit{blue}) $\Delta$BIC of a magnetic patch. The FOV of the inset matches the dimensions of the window described in Section~\ref{subsection:heightdependence}.}
\label{chi2_atmosmodels}
\end{figure*}

\subsection{Height-Dependent Magnetic Flux} \label{subsection:heightdependence}

In Figure~\ref{btotflux}, we report the total magnetic flux at optical depths along the LOS $\log_{10}\tau = -2.0$ to $0.0$. For the 1-component atmosphere the total flux with respect to optical depth ranges from $1.84 \times 10^{21}$~\si{Mx} up to $2.23 \times 10^{21}$~\si{Mx}, and a typical uncertainty is $0.04\times10^{21}$~\si{Mx}.

For the 2-component configuration, we report the results based on initiating the inversion with a filling factor of $f=0.5$. Introducing the filling factor overall reduces the flux estimates and increases its variation with height. The values now range from $1.01 \times 10^{21}$~\si{Mx} to $1.48 \times 10^{21}$~\si{Mx}, increasing with height (decreasing $\tau$). Here, the typical uncertainty is $0.03\times10^{21}$~\si{Mx}.

At $\log_{10} \tau = -1.5$, the 1-component flux is approximately $33.6\%$ higher than the 2-component flux. The total flux calculated from the 2-component SIR inversion, at $\log_{10}\tau = -1.5$, is approximately $28.2\%$ larger than the MERLIN inversion (dotted line in Figure~\ref{btotflux}). We provide a more detailed comparison with the MERLIN results and the literature in Section~\ref{subsec:literature}.

Figure~\ref{model_compare} compares the distribution of $f B_r$ (top row) and inclinations (bottom row) between the two configurations at five different optical depths. MERLIN quantities are included for reference, but note that the distribution is identical over the optical depth range since it is an ME inversion. In general, the 2-component configuration produces weaker $f B_r$ compared to the 1-component counterpart. Inclinations remain similar between the two configurations with a minimal increase in the outer inclinations closer to $0^\circ$ and $180^\circ$. This behavior is expected for the polar region, as the flux comes mainly from the transverse component of the field $B_t$, which scales with the filling factor as $f^{-0.5}$ \citep[e.g.,][]{pevtsov2021}.

\begin{figure}
\centering
\includegraphics[width=\columnwidth]{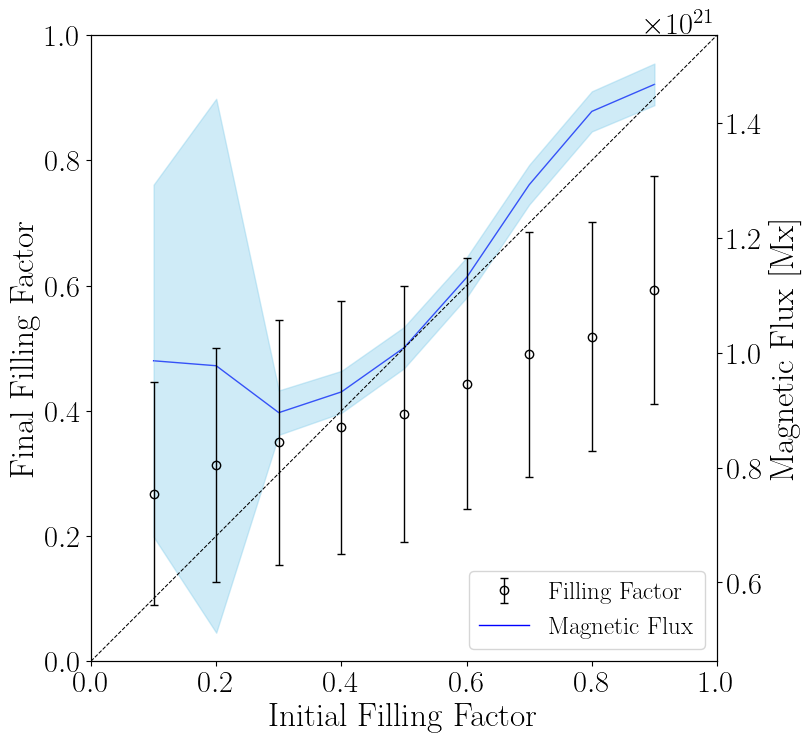}
\caption{Relationship between the initial guess and the output filling factors. The blue curve represents the magnetic flux at $\log_{10}\tau = -1.5$ for the respective initial filling factors.}
\label{init_fin_ff}
\end{figure}


\subsection{One vs Two-Component Configuration} \label{subsec:modeldegeneracy}

Figure~\ref{btotflux} shows that 1- vs. 2-component atmosphere models lead to significantly different total flux estimates. To assess the quality of the model, we plot the histograms of $\chi_\text{red}^2$ for the two cases in the left panel of Figure~\ref{chi2_atmosmodels}. The 2-component case has fewer pixels with $\chi_\text{red}^2>8$ compared to the 1-component case. The two distributions have a similar peak at about $4$, that is, the data noise is likely underestimated and/or the model is not complex enough to disentangle the two distributions.

We use the Bayesian Information Criterion \citep[BIC;][]{asensio2012,schwarz1978} to quantitatively compare the model results. BIC is calculated as 
\begin{equation}
    \textnormal{BIC} = \chi^2 + p \ln{\left(4N\right)},
    \label{bicequn}
\end{equation}
where $\chi^2=\left( 4N - p \right)\chi_\mathrm{red}^2$ is the raw chi-square (see Eqn.~\ref{chir}), $p$ is the number of free parameters, and $N$ is the total number of observed wavelengths. BIC is related to the $\chi^2$ computed in Section~\ref{sec:sir} but also penalizes excess dimensionality. 

The right panel of Figure~\ref{chi2_atmosmodels} shows the difference in the BIC values ($\Delta$BIC=BIC$_{\rm 2comp}$-BIC$_{\rm 1comp}$) between the 1- and 2-component models. The value of $\Delta$BIC is less than $-10$ for 58.3\% of the pixels, greater than $10$ for 32.3\%, and between $\pm10$ for 9.4\%. This suggests that the 2-component model is favored for a majority of the pixels (a strong model preference is considered for $|\Delta$BIC $|> 10$, \citealt{kass1995}).

The inset shows that the interiors of large magnetic patches are typically associated with negative $\Delta$BIC values, whereas the peripheries are where most of the positive $\Delta$BIC reside. Because the pixels at the peripheries typically have weaker polarization signals, further restricting our analysis to more stringent polarization selection criteria will result in a stronger overall preference for the 2-component model. We posit that the two configurations lead to similar, non-satisfactory fits of the Stokes profiles when the polarization is weak. As the 2-component configuration is penalized by its complexity, the BIC criterion will favor the 1-component configuration.


\subsection{Effect of Initial Value of $f$} \label{subsec:feffect}

For both 1- and 2-component schemes, different initial atmospheric models will lead to different $\chi^2$ and different $fB_r$ values. There is no single initial atmospheric model that works better than the rest (see Section~\ref{subsubsec:modelatmos}). The best initial model (the one that delivers the lowest $\chi^2$) differs from pixel to pixel with no discernible pattern, as shown in Figure~\ref{modelatmospheres} for a representative magnetic patch (the same as in Figure~\ref{polar_field_context}). The best initial model also differs between the 1- and 2-component inversion schemes.

Similarly, we find that the initial guess of $f$ affects the final result. We experiment with 9 different initial $f$ values ($f=0.1, 0.2, 0.3, ..., 0.9$) for the selected pixels in the full map. The black markers in Figure~\ref{init_fin_ff} show the relationship between the initial input and the final output filling factors. The average inverted filling factors positively correlate with initial guesses, but the slope is smaller than unity. The values of $f$ quickly drop below $0.6$ once $f$ is allowed to vary. The uncertainties are calculated from the standard deviation of the filling factors contained in the full map for each respective initial guess of $f$. Here, the minimum and maximum final filling factors are 0.27 and 0.59, respectively.

We calculate the total magnetic flux based on different initial guesses of the filling factor indicated by the secondary vertical axis in Figure~\ref{init_fin_ff} (blue curve). The average total flux ranges from $8.96 \times 10^{20}$~\si{Mx} to $1.47 \times 10^{21}$~\si{Mx}. The uncertainty of the flux is rather large for the smallest initial guesses of $f$, which suggests that the inversion likely converges to less satisfactory fits to the observation below the default of 0.5.

The fact that both the final filling factor and the magnetic flux estimate strongly depend on the initial guess of filling factor illustrates the ill-posed nature of the inversion problem for this observing setup. The minimization algorithm likely converges to one of many local minima in the $\chi^2$ space. In this case, the precision of the flux estimate is no better than a few tens of percent.


\begin{figure*}
\centering
\includegraphics[width=\textwidth]{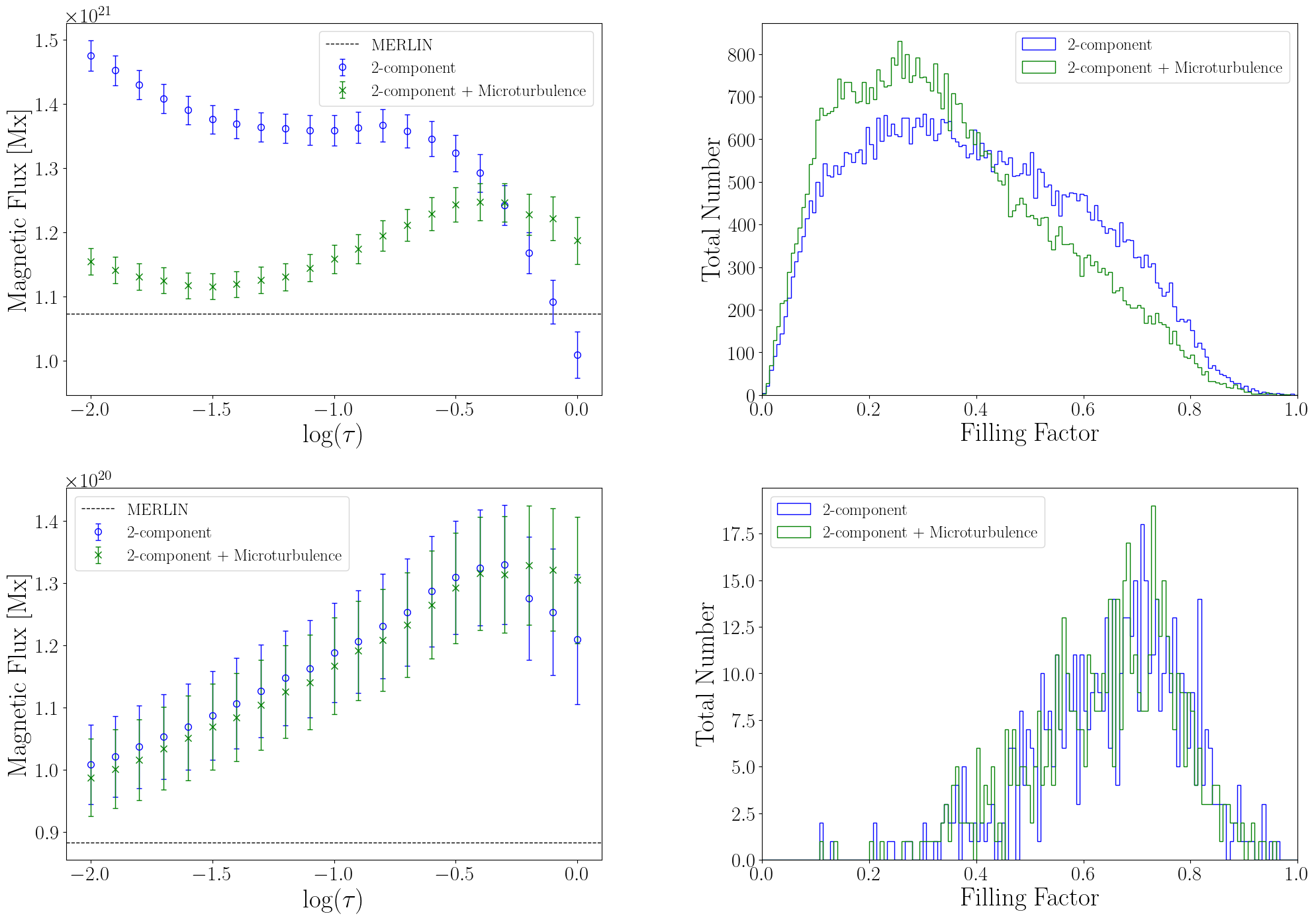}
\caption{Magnetic flux and filling factors for various inversion settings. The original 2-component configuration (blue) is shown against a modified 2-component configuration with an extra free parameter for microturbulence in the non-magnetic component (green). The dashed horizontal line represents the total flux calculated from the MERLIN inversion with the respective masks. \textit{Top Panels:} All significant pixels contained in the {full map mask}. \textit{Bottom Panels:} Pixels corresponding to the reduced mask, where the $Q$, $U$, and $V$ polarization signals are all above the selection threshold.}
\label{btotflux_degen_polsel}
\end{figure*}


\subsection{Effect of Inversion Settings} \label{subsec:effect2}

Inversion results are sensitive to how the atmosphere is composed, inclusion of parameters to invert, as well as the complexity of the atmosphere, specifically the amount of nodes for each parameter. To illustrate this point, we experiment with a modified 2-component configuration by including a height-independent microturbulence velocity as a free parameter for the non-magnetic atmosphere (Table~\ref{table:sirparams}). BIC analysis suggests that the new setting leads to an overall better fitting of the Stokes profiles. However, as shown in the top row of Figure~\ref{btotflux_degen_polsel}, it also leads to an overall lower magnetic flux above $\log_{10}\tau = -0.4$ and lower filling factors. At the nominal layer $\log_{10}\tau = -1.5$, the modified configuration yields a value of $(1.12 \pm 0.02) \times 10^{21}$~\si{Mx}, about $19\%$ lower than the original.

The inversion results can also be affected by noise. To this end, we impose a stricter pixel-selection criterion that Stokes $Q$, $U$, and $V$ must all exceed the threshold from Sec.~\ref{subsec:pixelselection}. This reduces our sample to 568 pixels, which will be referred to as the ``reduced mask." Although this filtering dramatically limits the number of usable pixels to $1\%$ of the {full map mask}, all pixels possess strong polarization signals and are expected to be less affected by noise. Visual inspection indicates that they reside mainly in the center of magnetic elements, as expected.

Two results are noteworthy.

First, pixels in this reduced mask consist mainly of vertical zenith angles, and \textit{none} are horizontal. At $\log_{10}\tau = -1.5$ for the 2-component (1-component) configurations, a total of $79.2\%\,(79.8\%)$ are vertical, $0.0\%\,(0.0\%)$ are horizontal, and $20.8\%\,(20.2\%)$ are undefined. This contrasts with the result for the {full map mask} presented in Section~\ref{subsec:disambiguation} and Figure~\ref{i1i2hist}.

Second, the discrepancies due to different inversion settings are greatly reduced. As shown in the bottom row of Figure~\ref{btotflux_degen_polsel}, the modified and nominal 2-component configurations now yield very similar results, in sharp contrast to the top row. Here, the magnetic flux smoothly decreases above $\log_{10}\tau = -0.3$. The distribution of the final filling factor peaks at about 0.7.

Our experiments suggest that for the strong-polarization pixels at the center of magnetic elements, the magnetic structure is consistent with a vertical flux tube gradually expanding with height. This also validates the radial-acute method for disambiguation. Nevertheless, when noisier pixels are included, various models become degenerate. Given the large degrees of freedom for the inversion settings, we again posit that the precision of our flux estimate is no better than several tens of percent.


\begin{figure}
\centering
\includegraphics[width=\columnwidth]{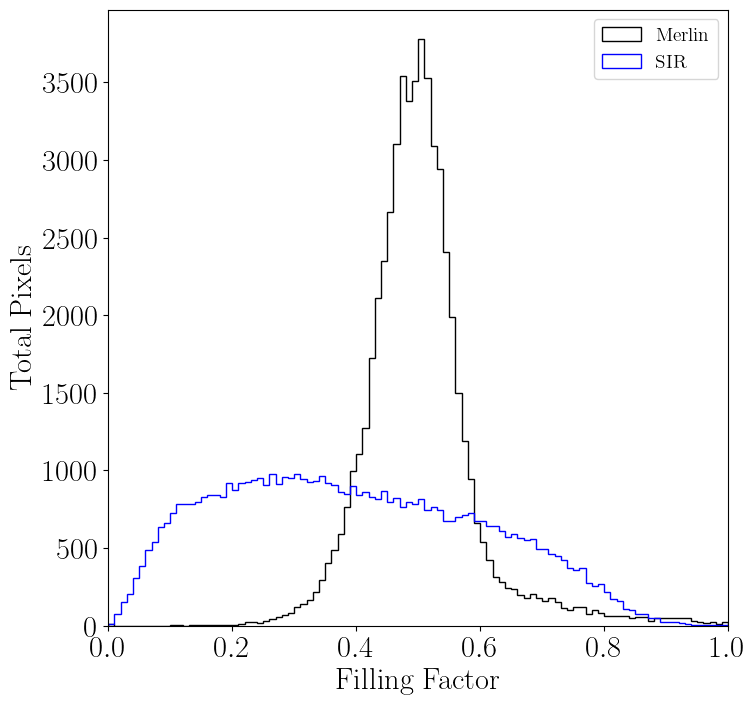}
\caption{Filling factors of pixels in the 2-component configuration for the SIR (blue) and MERLIN (black) inversions.}
\label{fillingfactor_1}
\end{figure}

\begin{figure*}
\centering
\includegraphics[width=\textwidth]{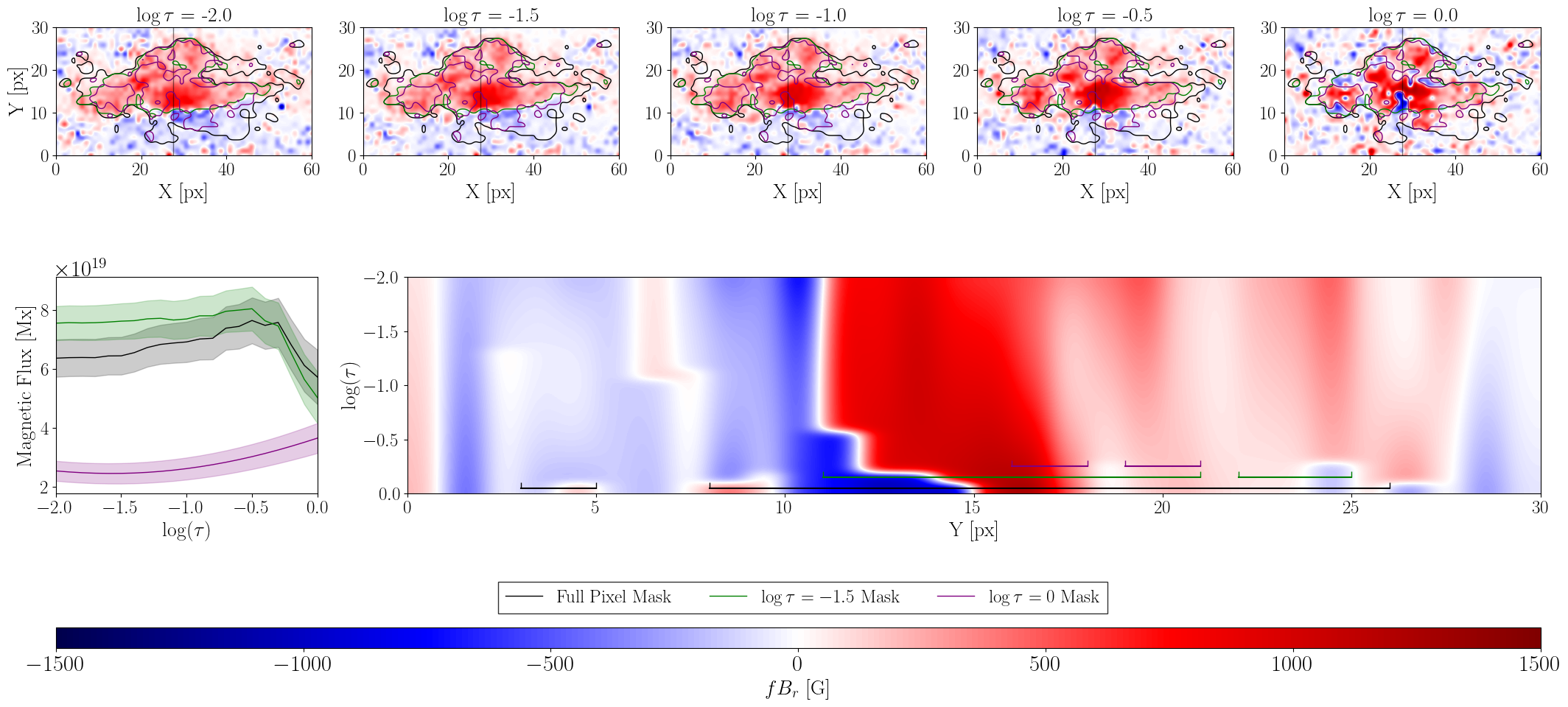}
\caption{Inversion results for the window FOV (60 by 30 pixels). \textit{Top row:} $fB_r$ at various optical depths. The black contours outline the pixels selected for our analysis. Two additional small window masks represent pixels where $fB_r$ is greater than \SI{100}{G} at $\log_{10}\tau = -1.5$ (green) and greater than \SI{100}{G} (purple) at $\log_{10}\tau = 0.0$. \textit{Bottom Left:} Total magnetic flux in the enclosed areas of the respective contours in the top panels. \textit{Bottom Right:} $fB_r$ slice corresponding to the vertical line position in the top panels. The bracket spans correspond to the masks indicated by the small window masks.}
\label{polarizationwindow}
\end{figure*}


\section{Discussion} \label{sec:discussion}


\subsection{Comparison with Previous Results} \label{subsec:literature}

For reference, we calculate the total magnetic flux using the publicly available MERLIN inversion results, which are based on the ME approximation and also include a filling factor free parameter. With the same pixel mask and disambiguation methods applied, we obtain a total flux of $1.07 \times 10^{21}$~\si{Mx} (Figure~\ref{btotflux}). The value roughly agrees with the SIR results at about $\log_{10}\tau=-0.1$, and is consistently about $20\%$ lower than the SIR results for layers above $\log_{10}\tau=-0.5$.

The final inverted filling factor is quite different in the SIR and MERLIN inversions. Figure~\ref{fillingfactor_1} shows that the final SIR filling factors (with an initial guess of 0.5) peak at approximately 0.27 and exhibit a broad full width at half maximum (FWHM) of 0.63. In comparison, the MERLIN filling factors peak at approximately 0.51 with a FWHM range of 0.13; the distribution is much narrower. However, this is a general comparison, where SIR uses an invertible non-magnetic model atmosphere (with 5 free parameters aside from the filling factor), while MERLIN specifies a stray light profile (fixed in shape) and only varies the filling factor and its wavelength shift in the inversion.

For the same data set, \cite{tsuneta2008} found a total vertical magnetic flux of $2.2 \times 10^{21}$~\si{Mx} when taking into account an effective filling factor. This flux is approximately $59.9\%$ higher than our flux at $\log_{10}\tau=-1.5$. \cite{tsuneta2008} determines the flux using a Milne-Eddington atmosphere with the MILOS inversion code \citep{orozcosuarez2007}, which does not allow for depth dependence. Additionally, MILOS uses a ``local" stray light profile, which is an intensity profile calculated as an average of the intensities surrounding the pixel under consideration. This typically leads to the inversion code to find much larger filling factors than MERLIN because the local stray light profile is similar to the intensity of the inverted pixel. While reconciliation of the two results is not possible without further analysis, we do note that our pixel selection criteria are more stringent. We speculate that the difference may be due in part to the different selection criteria of pixels: \cite{tsuneta2008} included a pixel total of 10.5\%, compared to our selection of 3.6\%. The additional pixels included in \citet{tsuneta2008} likely contain weaker polarization signals and thus weaker magnetic flux density compared to our sample.

Previous work shows that the polar magnetic flux is sensitive to the spatial resolution of the observation and can be affected by the systematics introduced with the inclusion of filling factor. \cite{leka2012} demonstrated the effects of resolution on high-resolution magnetic field data and additionally showed that $f$ is degenerate with other parameters. Such measurements are susceptible to systematic biases introduced during the inversion process, particularly for regions with low signal-to-noise ratios or extreme viewing angles, as is the case near the poles. These effects can influence both the retrieved field strengths and inclination distributions, and should be considered when interpreting our results \citep[see][]{leka2022}.

Studying a simulated solar atmosphere, \cite{milic2024} concluded that the magnetic flux density can be underestimated by more than $30\%$ when observed with a 20-cm telescope and the magnetic structures are unresolved. The bias is attributed to the non-linearity in the magnetic field inference process. A comprehensive analysis by \cite{sinjan2024} using analogous methods reached similar conclusions but went further: a significant fraction of the magnetic flux will be missed almost everywhere on the solar disk when the resolution is low and for extreme projection effects.

Our analysis provides another line of evidence that the flux estimate is influenced by various systematic effects. Each of the aforementioned scenarios can result in a difference of tens of percent. Unfortunately, more work is required to resolve the open flux problem.


\subsection{On the Height Dependence of Magnetic Flux} \label{subsec:height}

The SIR inversion allows us to explore the change of magnetic flux with height, but the trend is not always straightforward to interpret. For example, the magnetic flux appears to \emph{increase} slightly with height in the 2-component configuration for the {full map mask} (Figure~\ref{btotflux}). This contrasts with previous studies showing that the field strength typically declines with height as flux tubes expand, that is, the magnetic flux within a fixed mask should remain constant or decrease. The total $B_z$ must be constant with height $z$, unless field lines start leaving the mask. The loss of magnetic flux can be attributed to the optical depth mask, where the flux tubes will expand out of the mask, causing a loss in flux. \cite{solanki1999} modeled flux tube expansion and found a monotonic decrease in field strength with height due to flux conservation. \cite{pietarila2010} used \textit{Hinode} observations to show that network fields become more diffuse and bipolar higher in the atmosphere, consistent with canopy expansion and weaker fields aloft. Our experiment with the pixels with the strongest polarization signals in Section~\ref{subsec:effect2} also supports this view. Finally, the variation of flux with optical depth can be a consequence of the corrugation of the constant optical depth surfaces \citep[e.g.][]{milic2024}

Here we study a small window shown in Figure~\ref{polarizationwindow} in detail. Our nominal analysis uses the {full map mask} outlined by the black contour, which includes both positive and negative $fB_r$. We define two additional masks, an ``upper mask" for pixels with $fB_r>100$~G at $\log_{10}\tau=-1.5$ (green), and a ``lower mask" for $fB_r>100$~G at $\log_{10}\tau=0.0$ (purple). Both masks only include pixels with positive polarities at the $\tau$ layer where they are defined. They are designed to represent the conditions at different heights within the positive-polarity flux tube. Their relative locations are illustrated in the lower-right panel, on a vertical cross section.

The lower-left panel shows that the magnetic flux calculated for all three masks decrease gradually with height above $\log_{10}\tau=-0.4$. For the {full map mask} and the upper mask, there is an increase of flux in the lowest layers. These results yet again differ from Figure~\ref{btotflux}, but are illustrative of a couple factors that can affect the flux distribution with height.

First, the SIR inversion can well return greater $fB_r$ at higher layers. This is evident in the vertical view (lower right panel), where there are darker red portions closer to $\log_{10}\tau=-2$ compared to pixels in the lower layers in the same column (e.g., at $Y\approx13$, $19$, and $24$).

Second, the deeper layers may contain more minority polarities, which will affect the total flux. For this slice, the lower mask includes only positive polarity at all heights. The resultant flux thus decreases smoothly with height, as expected for simple magnetic flux tube expansion. The upper mask, defined at $\log_{10}\tau=-1.5$, contains only positive polarity at that height. However, it includes negative flux (e.g., at $11<Y<15$) in the lower layers, as the flux tube shrinks in area. The net flux thus becomes lower at these lower heights. The same argument applies for the {full map mask} in the $\log_{10}\tau<-0.4$ range.


\subsection{On the Analysis Scheme} \label{subsec:scheme}

Our specific choice of inversion settings may affect the results in unexpected ways. The following are several examples.
\begin{itemize}[parsep=0ex,partopsep=-0.5ex,itemsep=0.5ex,leftmargin=3mm]

\item \textit{Filling factor.}--- The filling factor is expected to decrease with height for expanding flux tubes. Unfortunately the SIR code does not allow for $\tau$-dependent filling factor at the moment.

\item \textit{Local nature of inversion.}--- Because the inversion treats each pixel in the FOV individually, the collective solution may or may not be physical when the global magnetic connectivity is taken into consideration. The unsigned magnetic flux, for example, is not required to be conserved.

\item \textit{Interpolation.}--- To obtain physical variables at certain optical depth, the SIR code interpolates between predetermined nodes, which spans the range $-4.2 < \log_{10}\tau < 1.0$. The lowest and highest nodes are located at the extrema of the $\log_{10}\tau$ space, where the inferred field is not well constrained (due to the small contribution function for the \ion{Fe}{1} 630~nm lines in higher layers). This may affect the interpolated results between the nodes.

\end{itemize}

Our radial-acute disambiguation scheme is expected to work well for the strong-polarization pixels at the center of magnetic elements. However, it is applied to each optical layer individually without considering neighboring pixels and optical-depth layers. This may induce inconsistent solutions. Furthermore, the disambiguation scheme strongly prefers vertical fields, which may not be appropriate for pixels with true horizontal fields. Inclined, open magnetic fields at the periphery of magnetic elements can still contribute to magnetic flux. Inclined closed fields at the foot points of low-lying loops can strongly affect the coronal dynamics through reconnection with open fields. By omitting pixels with weaker linear polarization, our result of the magnetic flux may be overestimated, and our view of the polar magnetism will be incomplete.


\section{Summary and Outlook}

In this work, we analyze a raster map of the southern polar coronal hole observed by \textit{Hinode}/SP. Inversions using the SIR code provide height-dependent vector magnetic field maps between optical depths $\log_{10}\tau = -2$ and $0$. For optical depth $\log_{10}\tau = -1.5$, we estimate the total magnetic flux to be $(1.84 \pm 0.03) \times 10^{21}$~\si{Mx} and $(1.38 \pm 0.02) \times 10^{21}$~\si{Mx} under the 1- and 2-component atmosphere configurations, respectively. The magnetic flux is approximately constant or increases slightly with height in the two cases, respectively. The 2-component result at $\log_{10}\tau = -1.5$ is about 28.2\% greater of the MERLIN inversion result.

The inversions allow us to examine the impact on the magnetic flux estimate from adopting (1) 1- vs 2-component atmospheric models via filling factor, and (2) different analysis schemes. The BIC test shows that most pixels favor the 2-component configuration, but a non-negligible portion still favors the 1-component configuration. The optimal estimate of the flux is likely somewhere in between the configurations. In addition, the results also depend on the initial input atmosphere models, initial filling factor, and inversion settings. In cases where the results are sensitive to these factors, the minimization scheme likely falls into different local minima in the $\chi_\text{red}^2$ space. These systematic effects severely limit the precision of the polar magnetic flux inferences. 
 
While our analysis focuses on the effects of model degeneracies and the different magnetic inferences depending on the initial guess, these issues are particularly relevant in observations of the polar regions (and near-limb observations in general). This is because foreshortening strongly reduces spatial resolution, and the quadratic dependence of the linear polarization with the transverse field (responsible for magnetic flux) further complicates magnetic field inferences. Our results suggest that common inversion approaches may lead to serious systematics that are hitherto largely neglected. This has direct implications for the open flux problem, as inferred photospheric magnetic field measurements that are used as boundary conditions in coronal extrapolations may be biased and will contribute to the mismatch with \textit{in-situ} heliospheric flux measurements. However, our results show that inconsistent inversion inferences are reconciled when the signals are well above the noise.

The limitations discussed can be investigated with data from the \textit{Daniel K. Inouye Solar Telescope} \citep[\textit{DKIST};][]{rimmele2020}. Its extremely high spatial/spectral resolution (sub $0\farcs1$) and simultaneous multi-line diagnostics (probing different atmospheric layers) allow for resolving the fine-scale structures of polar magnetic patches while simultaneously providing height stratification. As suggested by previous numerical experiments \citep{centeno2023,milic2024,sinjan2024}, \textit{DKIST} observations may provide an answer as to whether the open flux problem is caused by the limited resolution.

In addition to higher-resolution data, more advanced analysis techniques are another possibility to alleviate the limitations. Implementations such as spatially coupled inversions \citep{vannoort2012,danilovic2016,delacruzrodriguez2019} aim to properly account for image degradation effects that are generally approximated or not considered. They utilized spectral and spatial information to parametrize the atmosphere in the FOV simultaneously, minimizing the errors introduced by excessive free parameters. This directly reduces the complexity of 2-component configurations as considered in our study. Disambiguation methods that seek globally optimized solutions, such as the (fine-tuned) Minimum-Energy method or inversions directly coupled with disambiguation \citep[e.g.,][]{borrero2019,pastoryabar2019}, can also help by providing additional constraints that are more physically motivated.


\section*{Acknowledgments} %
Funding for the \textit{DKIST} Ambassadors program is provided by the National Solar Observatory, a facility of the National Science Foundation, operated under Cooperative Support Agreement number AST-1400450. B.Y., X.S., I.M., and M.G. are additionally supported by NASA HGIO award 80NSSC21K1283. \textit{Hinode} is a Japanese mission developed and launched by ISAS/JAXA, with NAOJ as domestic partner and NASA and STFC (UK) as international partners. It is operated by these agencies in co-operation with ESA and NSC (Norway).
R.C. acknowledges support from NASA LWS Award 80NSSC20K0217. This material is based upon work supported by the NSF National Center for Atmospheric Research, which is a major facility sponsored by the U.S. National Science Foundation under Cooperative Agreement No. 1852977. C.Q.N. acknowledges support from the Agencia Estatal de Investigaci\'{o}n del Ministerio de Ciencia, Innovaci\'{o}n y Universidades (MCIU/AEI) under grant ``Polarimetric Inference of Magnetic Fields", the European Regional Development Fund (ERDF) with reference PID2022-136563NB-I00/10.13039/501100011033 and the Project ICTS2022-007828, funded by MICIN and the European Union NextGenerationEU/RTRP. A.P.Y. acknowledges support from the Swedish Research Council (grant 2023-03313). This project has been funded by the European Union through the European Research Council (ERC) under the Horizon Europe program (MAGHEAT, grant agreement 101088184). M.G. acknowledges support from NASA contract NNM07AA01C (Solar-B (Hinode) Focal Plane Package Phase E).

The technical support and advanced computing resources from University of Hawaii Information Technology Services – Research Cyberinfrastructure, funded in part by the National Science Foundation CC* awards $\#2201428$ and $\#2232862$ are gratefully acknowledged.

We thank the anonymous referee for their exceptionally thorough and constructive review, whose detailed feedback and literature recommendations led us to reanalyze the data, repeat all spectropolarimetric inversions, and achieve a more comprehensive and robust set of results.

\facilities{\textit{Hinode}} 


\setcounter{section}{0} 
\renewcommand{\thesection}{\Alph{section}} 

\section*{Appendix}
\section{Disambiguation Methods} \label{sec:appendix}

The 180$^\circ$ azimuth ambiguity in spectropolarimetric observations is commonly resolved using ``Minimum Energy" algorithms \citep{metcalf1994}. These methods select the azimuthal direction that minimizes the divergence of the field ($\nabla \cdot \mathbf{B}$) and the electric current density ($J_z$). Publicly available disambiguation codes, such as ME0\footnote{\url{https://www.cora.nwra.com/AMBIG/ME0.html}} are frequently applied to \textit{Hinode}/SP data for this purpose. We have applied the code with its default parameters, unfortunately the results are unsatisfactory.

In coronal holes, magnetic fields in strong-polarization elements are expected to be approximately radial \citep[e.g.,][]{petrie2009}. At high heliographic latitudes, the LOS direction becomes highly inclined relative to the local vertical, so that the LOS component no longer reliably represents the vertical magnetic field. This may affect the performance of ME0, which relies on a potential field extrapolation based on $B_{\mathrm{LOS}}$ to calculate $\nabla \cdot \mathbf{B}$. The method is also sensitive to both projection effects and transverse field uncertainties, which can significantly degrade ambiguity resolution performance away from the disk center \cite{leka2009}.

We have also experimented the geometry-based method following \citet{ito2010}. To this end, we calculate the two zenith angles $i_{1,2}$ for the two possible azimuth solutions $\psi_{1,2}$. The authors introduce an assumption that the magnetic field vector is either vertical (zenith angle $0^\circ - 40^\circ$, or $140^\circ - 180^\circ$) or horizontal ($70^\circ - 110^\circ$) to the local surface, or undefined. If both $i_1$ and $i_2$ solutions are vertical, then the inclination closer to the local normal is selected. If one solution is vertical and the other horizontal, the correct solution is ambiguous and the pixel is not considered. Lastly, if one solution is vertical or horizontal and the other is undefined, then the field will adopt the one with the defined solution.

We find that a significant fraction of the pixels fall within the range of undefined, as qualitatively illustrated in the bottom row of Figure~\ref{i1i2hist}. As they are excluded from analysis, the resultant magnetic flux is much smaller compared to our nominal result or the literature.

\renewcommand{\thesection}{\arabic{section}}


 
\bibliographystyle{aasjournal}
\bibliography{biblio.bib}

\end{CJK*}

\end{document}